\documentclass[twocolumn,jcp,aip,showpacs,superscriptaddress,amssymb,amsmath,reprint]{revtex4-1}
\usepackage{graphicx}
\usepackage{epstopdf}
\usepackage{bm}
\usepackage{hyperref}
\usepackage{comment}

\hypersetup{%
   pdfpagemode=None, 
   pdfstartpage=1,
   pdfmenubar=true,
   pdftoolbar=true,
   colorlinks = true,
   linkcolor=blue,
   citecolor=blue,
   urlcolor=blue,
   bookmarksopen=false
 }

\newcommand{\be}{\begin{equation}}
\newcommand{\ee}{\end{equation}}

\begin{document}
\title{Interactions of benzene, naphthalene, and azulene with alkali-metal\\ and alkaline-earth-metal atoms for ultracold studies}
\author{Pawe\l~W\'ojcik}
\affiliation{Faculty of Physics, University of Warsaw, Pasteura 5, 02-093 Warsaw, Poland}
\author{Tatiana~Korona}
\affiliation{Faculty of Chemistry, University of Warsaw, Pasteura 1, 02-093 Warsaw, Poland}
\author{Micha\l~Tomza}
\email{michal.tomza@fuw.edu.pl}
\affiliation{Faculty of Physics, University of Warsaw, Pasteura 5, 02-093 Warsaw, Poland}
   
\date{\today}

\begin{abstract}

We consider collisional properties of polyatomic aromatic hydrocarbon molecules immersed into ultracold atomic gases and investigate intermolecular interactions of exemplary benzene, naphthalene, and azulene with alkali-metal (Li, Na, K, Rb, Cs) and alkaline-earth-metal (Mg, Ca, Sr, Ba) atoms. We apply the state-of-the-art \textit{ab initio} techniques to compute the potential energy surfaces (PESs). We use the coupled cluster method restricted to single, double, and noniterative triple excitations to reproduce the correlation energy and the small-core energy-consistent pseudopotentials to model the scalar relativistic effects in heavier metal atoms. We also report the leading long-range isotropic and anisotropic dispersion and induction interaction coefficients. The PESs are characterized in detail and the nature of intermolecular interactions is analyzed and benchmarked using symmetry-adapted perturbation theory. The full three-dimensional PESs are provided for selected systems within the atom-bond pairwise additive representation and can be employed in scattering calculations. Presented study of the electronic structure is the first step towards the evaluation of prospects for sympathetic cooling of polyatomic aromatic molecules with ultracold atoms. We suggest azulene, an isomer of naphthalene which possesses a significant permanent electric dipole moment and optical transitions in the visible range, as a promising candidate for electric field manipulation and buffer-gas or sympathetic cooling.

\end{abstract}

\maketitle

\section{Introduction}

Intermolecular interactions are essential in many areas of natural sciences because they govern properties and dynamics of molecular systems at the microscopic level in phenomena ranging from folding proteins and photosynthetic light harvesting in biology to chemical reactions and self-organization of nanostructures in solid-state physics~\cite{Stone13}. Experiments at low and ultralow temperatures provide a useful playground for answering questions touching upon the fundamentals of quantum mechanics in a controlled and systematic way~\cite{BlochRMP08}. At ultracold conditions even a tiny change in the interaction energy can be larger than the collision energy and thus can modify the rates of elastic, inelastic, and chemically reactive scattering by many orders of magnitude~\cite{QuemenerCR12}. Therefore, a combination of experimental and theoretical efforts applied to study molecules at ultralow temperatures can be very instructive and can shed new light on intermolecular interactions~\cite{ShagamNatChem15,KleinNatPhys16,TomzaPRL15}. 

The first spectacular successes in the field of ultracold quantum matter were achieved with atoms~\cite{BlochRMP08}.
However, molecules have additional rotational and vibrational degrees of freedom that could potentially be used for various applications~\cite{CarrNJP09,DulieuPCCP11}. Therefore, diatomic alkali-metal molecules were produced in their absolute rovibrational ground state~\cite{NiScience08} and employed in a series of groundbreaking experiments on controlled chemical reactions~\cite{OspelkausScience10,NiNature10,MirandaNatPhys11} and quantum simulations~\cite{BoNature13}. Fueled by the promise of exciting applications~\cite{CarrNJP09}, the production of more complex and polyatomic molecules at ultralow temperatures is currently emerging as another important research goal.

Recently, the first experiments on cooling of polyatomic molecules have been launched. Ammonia (NH$_3$)~\cite{BethlemNature00} and methyl radical (CH$_3$)~\cite{LiuPRL17} were cooled down to low ($<1\,$K) temperatures with Stark and Zeeman decelerators and subsequently trapped in electric and magnetic traps, respectively. Cold fluoromethane (CH$_3$F) was produced using centrifuge decelerator~\cite{WuScience17}. Fluoromethane (CH$_3$F)~\cite{ZeppenfeldNature12}, formaldehyde (H$_2$CO)~\cite{GlocknerPRL15,PrehnPRL16}, and strontium monohydroxide (SrOH)~\cite{KozyryevPRL17a,KozyryevPRL18} were successfully cooled down to ultralow ($<1\,$mK) temperatures using Sisyphus laser cooling. Such laser-cooled polyatomic molecules can find applications in precision measurement of the time-reversal symmetry violation~\cite{KozyryevPRL17} and the time variation of the proton-to-electron mass ratio~\cite{Kozyryev2018}. Laser cooling of complex polyatomic molecules with six or more atoms was also theoretically proposed~\cite{KozyryevCPC16,IsaevPRL16,Kozyryev2018b,ORourke2019} and loading polyatomic molecules into a magneto-optical trap is the expected next step. Helium buffer-gas cooling of benzonitrile (C$_6$H$_6$CN)~\cite{PattersonPCCP15} and \textit{trans}-stilbene (C$_{14}$H$_{12}$)~\cite{PiskorskiCPC14} to low temperatures were demonstrated and opened the way for slowing down and trapping of polyatomic aromatic molecules at ultralow temperatures.   

However, to produce molecules at even lower temperatures, of the order of $\mu$K, a second-stage cooling process is required.
One promising technique is sympathetic cooling in which temperature of pre-cooled molecules is further reduced by thermal collisional contact with much colder ultracold atomic gas. Prospects for sympathetic cooling of diatomic molecules with alkali-metal or alkaline-earth-metal atoms have been theoretically investigated for several systems (see e.g.~Refs.~\cite{SoldanPRL04,LaraPRL06,WallisPRL09,MoritaPRA18}) but just a few works have considered larger molecules~\cite{ZuchowskiPRA08,ZuchowskiPRA09,TscherbulPRL11,LoreauJCP15,Morita2017} and experimentally only ammonia molecules were immersed into ultracold rubidium atoms~\cite{ParazzoliPRL11}. Cold collisions and sympathetic cooling of molecules as large as benzene were theoretically investigated only for mixtures with helium and other rare-gas atoms~\cite{BarlettaNJP09,LiJCP12,CuiJCP14,LiJCP14,CroftPRA15,CuiJCP15}.
Intermolecular interactions of benzene and naphthalene with rare-gas atoms were investigated theoretically and experimentally~\cite{WeberCPL91,HobzaJCP92,KlopperJCP94,BrupbacherJCP94,RiedleJCP96, KochJCP98,KochJCP99,PiraniCPL01,ClementiJPCA01,CappellettiJPCA02,PiraniCPL03,CapeloJPCA09,MakarewiczJCP11,CalvoJCP13,CybulskiJPCA14,ShirkovJCP15}. Unfortunately, there is very limited knowledge of cold interactions and collisions between large polyatomic molecules and alkali-metal or alkaline-earth-metal atoms, hence prospects for sympathetic cooling of such molecules down to low and ultralow temperatures are not known.

In the present work, we investigate intermolecular interactions of three representative polyatomic aromatic hydrocarbon molecules, i.e.~benzene, naphthalene, and azulene, with alkali-metal (Li, Na, K, Rb, Cs) and alkaline-earth-metal (Mg, Ca, Sr, Ba) atoms using state-of-the-art \textit{ab initio} methods of quantum chemistry. Intermolecular interactions in this class of systems have not yet been extensively studied (see e.g.~Refs.~\cite{BakerJPCA10,DenisCPL13,DenisCP14,SadlejPCCP15,BorcaJPCA16,UllahCPL18}), especially in the context of cold experiments, and here we fill this gap. We calculate and characterize in detail the potential energy surfaces (PESs) and long-range dispersion and induction interaction coefficients. We analyze the nature of intermolecular interactions using the symmetry adapted perturbation theory (SAPT)~\cite{JeziorskiCR94} and benchmark this method in SAPT(HF) and SAPT(DFT)~\cite{Misquitta:05,HesselmannJCP05} variants while applied to aromatic hydrocarbons interacting with metal atoms. Finally, we consider consequences of our findings and prospects for sympathetic cooling of polyatomic aromatic molecules with ultracold atoms.

The plan of this paper is as follows. Section~\ref{sec:theory} describes the theoretical methods used in the \textit{ab initio} electronic structure calculations. Section~\ref{sec:results} presents and discusses the intermolecular interactions of benzene, naphthalene, and azulene with alkali-metal and alkaline-earth-metal atoms. Section~\ref{sec:summary} summarizes our paper and discusses future possible applications.

\section{Computational details}
\label{sec:theory}

Benzene (C$_6$H$_6$), naphthalene (C$_{10}$H$_8$), and azulene (C$_{10}$H$_8$) are very stable closed-shell polyatomic molecules with aromatic bonds of delocalized $\pi$ electrons, which determine their properties, including rigid planar geometries~\cite{KrygowskiCR01}. These aromatic molecules are chemically stable while interacting with alkali-metal and alkaline-earth-metal atoms. Interactions in such systems are of non-covalent nature dominated by the dispersion and induction contributions. Thus, their electronic ground state inherits the doublet and singlet spin symmetry of alkali-metal and alkaline-earth-metal atoms, respectively. Benzene and naphthalene are apolar, whereas azulene posseses a significant permanent electric dipole moment of around 0.8$\,$Debye~\cite{ToblerJMS65}. We describe these molecules within the rigid rotor approximation assuming their geometrical structures accurately determined by high-resolution spectroscopy~\cite{BabaJCP11,HuberMP05}.

In order to investigate intermolecular interactions, we adopt the computational scheme successfully applied to the ground-state interactions between polar alkali-metal dimer~\cite{TomzaPRA13b} and polyatomic molecular ions with alkali-metal and alkaline-earth-metal atoms~\cite{TomzaPCCP17}. 
Thus, to calculate PESs for molecules interacting with alkaline-earth-metal atoms (alkali-metal atoms) we employ the closed-shell (spin-restricted open-shell) coupled cluster method restricted to single, double, and noniterative triple excitations, starting from the restricted closed-shell (open-shell) Hartree-Fock orbitals, CCSD(T)~\cite{PurvisJCP82,KnowlesJCP93}. The interaction energies are obtained with the supermolecular method and
the basis set superposition error is corrected by using the counterpoise correction~\cite{BoysMP70}
\begin{equation}\label{eq:intenergy}
E_\mathrm{int}=E_\mathrm{mol+at}-E_\mathrm{mol}-E_\mathrm{at}\,,
\end{equation}
where $E_\textrm{mol+at}$ denotes the total energy of the molecule interacting with the atom, and $E_\textrm{mol}$ and $E_\textrm{at}$ are the total energies of the molecule and atom computed in the dimer basis set. Calculations are carried out for around 25-35 intermolecular distances in the range of 3-30$\,$bohr.

The Li, Na, and Mg atoms are described with the augmented correlation-consistent polarized core-valence quadruple-$\zeta$ quality basis sets (aug-cc-pCVQZ)~\cite{PrascherTCA10}, whereas the H and C atoms are described with the augmented correlation-consistent polarized valence triple-$\zeta$ quality basis sets (aug-cc-pVTZ)~\cite{DunningJCP89,KendallJCP92}. The higher quality basis sets are used for metal atoms to account for their larger polarizabilities and smaller binding energies of valence electrons. The scalar relativistic effects in K, Rb, Cs, Ca, Sr and Ba atoms are included by employing the small-core relativistic energy-consistent pseudopotentials (ECP) to replace the inner-shells electrons~\cite{DolgCR12}. The use of the pseudopotentials allows one to use larger basis sets to describe the valence electrons and models the inner-shells electrons density as accurately as the high quality atomic calculation used to fit the pseudopotentials.
The pseudopotentials from the Stuttgart library are employed in all calculations. 
The K, Ca, Rb, Sr, Cs, and Ba atoms are  described with the ECP10MDF, ECP10MDF, ECP28MDF, ECP28MDF, ECP46MDF, and ECP46MDF pseudopotentials~\cite{LimJCP06,DolgTCA98} and the $[11s11p5d3f]$, $[12s12p7d4f2g]$, $[14s14p7d6f1g]$, $[14s11p6d5f4g]$, $[12s11p6d4f2g]$, and $[13s12p6d5f4g]$ basis sets, respectively, obtained by decontracting and augmenting the basis sets suggested in Refs.~\cite{LimJCP06,DolgTCA98}. 
The used basis sets were optimized in Refs.~\cite{TomzaPRA13a,TomzaPRA14,TomzaPRA15}.
The basis sets are additionally augmented in all calculations by the set of the $[3s3p2d]$ bond functions~\cite{midbond}.

To analyze the nature of intermolecular interactions, we employ the symmetry adapted perturbation theory (SAPT)~\cite{JeziorskiCR94}, which allows to decompose the interaction energy into the series of different contributions 
\begin{equation}
\begin{split}
E_\mathrm{int}=&E^{(1)}_\mathrm{elst}+E^{(1)}_\mathrm{exch}+E^{(2)}_\mathrm{disp}+\\
               &E^{(2)}_\mathrm{ind}+E^{(2)}_\mathrm{exch-disp}+E^{(2)}_\mathrm{exch-ind}+\dots\,,
\end{split}
\end{equation}
where $E^{(1)}_\mathrm{elst}$ and $E^{(1)}_\mathrm{exch}$ are the first-order electrostatic and exchange energies, $E^{(2)}_\mathrm{disp}$ and $E^{(2)}_\mathrm{ind}$ are the second-order dispersion and induction energies, and $E^{(2)}_\mathrm{exch-disp}$ and $E^{(2)}_\mathrm{exch-ind}$ are the second-order exchange-dispersion and exchange-induction energies.

We compute the SAPT interaction energies using two variants of the monomer description: the Hartree-Fock method (SAPT(HF))~\cite{JeziorskiCR94} and the density functional theory (SAPT(DFT))~\cite{Misquitta:05,HesselmannJCP05}. The PBE0 functional~\cite{Perdew:96,Adamo:99} with the asymptotic correction~\cite{Gruening:01} is used in calculations on the DFT level. The SAPT interaction energies are corrected by applying the Hartree-Fock delta correction~\cite{Moszynski:hartree}
\begin{equation}
\delta_\mathrm{HF} = E_\mathrm{int}^\mathrm{HF} -  \left( E^{(1)}_\mathrm{elst} + E^{(1)}_\mathrm{exch} + E^{(2)}_\mathrm{ind} + E^{(2)}_\mathrm{exch-ind}\right),
\end{equation}
where $E_\mathrm{int}^\mathrm{HF}$ is the Hartree-Fock supermolecular interaction energy as defined in Eq.~\eqref{eq:intenergy} and the SAPT components are calculated within SAPT(HF). With this correction the SAPT interaction energy is defined as 
\begin{equation}
E^{\mathrm{SAPT}+\delta_\text{HF}}_\mathrm{int} = E^\mathrm{(12)}_\mathrm{tot} + \delta_\mathrm{HF},
\end{equation}
where $E^\mathrm{(12)}_\mathrm{tot}$ is a sum of all SAPT terms of the first and of the second order. 

\begin{figure}[t!]
\begin{center}
\includegraphics[width=0.8\columnwidth]{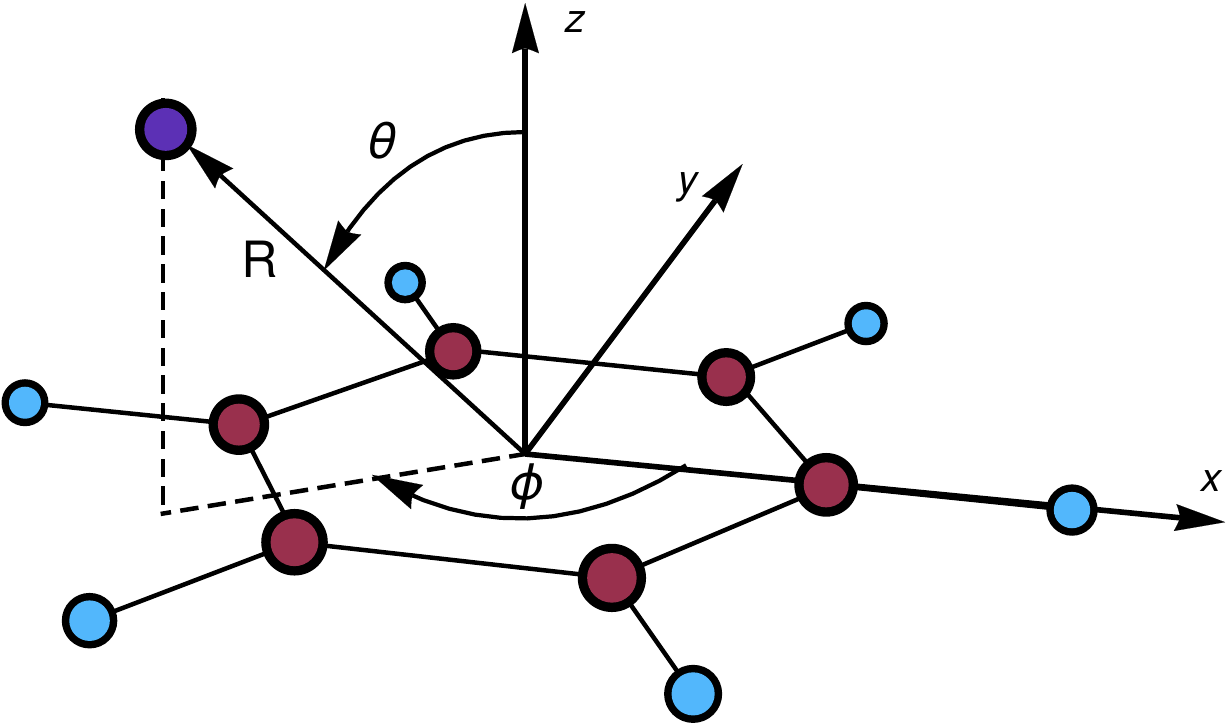}
\end{center}
\caption{The coordinates used to describe the benzene-metal complexes: $R$ is the relative distance between centers of mass, $\theta$ is the angle between $R$ and the 6-fold symmetry axis of benzene, and $\phi$ is the angle between the projection of $R$ onto the molecular plane and the axis, which is parallel to a C–H bond of benzene.}
\label{fig:geom}
\end{figure}

Potential energy surfaces for investigated systems within the rigid rotor approximation are three-dimensional functions $E_\text{int}(R,\theta,\phi)$ (see Fig.~\ref{fig:geom}). Their analytical forms may be useful for scattering calculations, therefore we provide the force fields (FFs) within the atom-bond pairwise additive representation~\cite{PiraniAtomBond}. In this model, the analytical form of PES is represented as a sum of interactions between an atom and every bond of the molecule. Here, we slightly modify the representation suggested in Ref.~\cite{PiraniAtomBond}. 

The atom-bond potential $V_{ab}(r,\theta)$ depends on the orientation of the bond and is represented as a linear combination of two one-dimensional potentials
\begin{equation}
\label{eq:lk}
V_{ab}(r,\theta) = V_{ab}^\parallel(r) \cos^2\theta + V_{ab}^\perp(r) \sin^2\theta\,,
\end{equation}
where $r$ is a distance between the atom and the geometric center of the bond while $\theta$ is an angle between the axis of the bond and the axis
connecting the atom with the center of the bond. One-dimensional potentials are polynomial functions
\begin{equation}
\label{eq:V1d}
V_{ab}^k(r) = \epsilon^k_{ab} \left(\frac{m}{\beta-m}\left(\frac{r^k_{ab}}{r}\right)^\beta - \frac{\beta}{\beta-m}\left(\frac{r^k_{ab}}{r}\right)^m \right)\,,
\end{equation} 
where $r^k_{ab}$ and $\epsilon^k_{ab}$ are parameters, which can be interpreted as the well depth and inter-species equilibrium distance in the atom-bond interaction model and which are different for parallel ($k=\parallel$) and perpendicular ($k=\perp$) components, as well as for different bond types: carbon-carbon ($ab$=CC) and carbon-hydrogen ($ab$=CH). $\beta$ and $m$ are constants of the model which describe behavior of the short-range repulsion and long-range attraction, respectively. The value of $m$ is set to 6 which is the scaling of the long-range dispersive interaction between two neutral species. The value of $\beta$ is chosen to be equal to 8 which is the suggested value for benzene - soft neutral atom interaction~\cite{PiraniAtomBond}. Numerical tests confirm that $\beta = 8$ assures the best performance of the force field. 

The interaction energy given by the force field is a sum of all atom-bond potentials present in the system
\begin{equation}
\label{eq:ff}
E_\text{int}^\text{FF}(R,\theta,\phi) = \sum_{ab } V_{ab}(r_{ab},\theta_{ab})\,.
\end{equation}
Values of parameters in our model are obtained by numerical minimization of the absolute difference between force field and \textit{ab initio} values for a set of calculated points, $\chi = \sum_i \left| E_\text{int}^\text{FF}(R_i,\theta_i,\phi_i) - E_\mathrm{int}(R_i,\theta_i,\phi_i)\right|$, where around 20 intermediate distances are selected avoiding too large short-range repulsive and too small long-range values.
We find the optimal parameters by running an extensive Monte Carlo search followed by local optimizations.

\begin{table*}[bth!]
\caption{Characteristics of the PESs for benzene interacting with alkali-metal and alkaline-earth-metal atoms, all in the ground electronic state: the equilibrium intermolecular distance $R_e$ and well depth $D_e$ for the global minima at the out-of-plane geometry, and saddle points at the side-in-plane geometry (with prime) and at the vertex-in-plane geometry (with double prime).\label{tab:spec}} 
\begin{ruledtabular}
\begin{tabular}{lrrrrrr}
System & $R_e\,$(bohr) & $D_e\,$(cm$^{-1}$) & $R_e'\,$(bohr) & $D_e'\,$(cm$^{-1}$) & $R_e''\,$(bohr) & $D_e''\,$(cm$^{-1}$) \\
\hline
C$_6$H$_6$+Li & 4.25 & 1500 & 11.2 & 145 & 11.9 & 109 \\
C$_6$H$_6$+Na & 6.05 & 756 & 11.5 & 141  & 12.2 & 107\\
C$_6$H$_6$+K &  6.17 & 1050 & 12.3 & 136 & 12.9 & 109\\
C$_6$H$_6$+Rb & 6.42 & 1100 & 12.5 & 138 & 13.1 & 111 \\
C$_6$H$_6$+Cs & 6.65 & 1280 & 12.8 & 138 & 13.4 & 113\\
C$_6$H$_6$+Mg & 6.77 & 758 & 10.7 & 219 & 11.5 & 148\\
C$_6$H$_6$+Ca & 6.86 & 960 & 11.6 & 220 & 12.3 & 159\\
C$_6$H$_6$+Sr & 6.73 & 1090 & 11.9 & 222 & 12.6 & 164\\
C$_6$H$_6$+Ba & 6.20 & 1640 & 12.3 & 220 & 13.0 & 168\\
\end{tabular}
\label{tab:spec}
\end{ruledtabular}
\end{table*}

Long-range interactions are important for studies of cold and ultracold collisions. The leading part of the intermolecular interaction energy between a closed-shell symmetric-top molecule and a $S$-state atom, both in the electronic ground state, at large intermolecular distances $R$, in the molecular frame, is of the form
\begin{equation}\label{eq:long-range}
E_\text{int}(R,\theta,\phi)\approx-\frac{C_{6,0}}{R^6}-\frac{C_{6,2}}{R^6}P_2(\cos\theta)+\dots\,,
\end{equation}  
where $C_{6,0}$ and $C_{6,2}$ are leading long-range isotropic and anisotropic interaction coefficients. For apolar molecules, they are given by the dispersion interaction only
\begin{equation}
\begin{split}
C^\mathrm{disp}_{6,0}&=\frac{3}{\pi}\int_0^\infty\alpha_\mathrm{at}(i\omega){\bar\alpha}_\mathrm{mol}(i\omega)d\omega\\
C^\mathrm{disp}_{6,2}&=\frac{1}{\pi}\int_0^\infty\alpha_\mathrm{at}(i\omega){\Delta\alpha}_\mathrm{mol}(i\omega)d\omega\,,
\end{split}
\end{equation}
where ${\alpha}_{\textrm{atom(molecule)}}(i\omega)$ is the dynamic polarizability of the atom(molecule) at imaginary frequency and the
average polarizability and polarizability anisotropy are given by ${\bar\alpha}=(\alpha_{xx}+\alpha_{yy}+\alpha_{zz})/3$ and ${\Delta\alpha}=\alpha_{zz}-\frac{\alpha_{xx}+\alpha_{yy}}{2}$, respectively. For polar molecules, both dispersion and induction interactions contribute to the long-range interaction coefficients, $C_{6,0}=C^\mathrm{disp}_{6,0}+C^\mathrm{ind}_{6,0}$ and $C_{6,2}=C^\mathrm{disp}_{6,2}+C^\mathrm{ind}_{6,2}$, with
\begin{equation}
C^{\mathrm{ind}}_{6,0}=C^{\mathrm{ind}}_{6,2}=d^2_\textrm{mol}\alpha_\textrm{atom} \,,\\
\end{equation}
where $d_\textrm{mol}$ is the permanent electric dipole moment of the molecule and $\alpha_\textrm{atom}$ is the static electric dipole polarizability of the atom.

The dynamic electric dipole polarizabilities at imaginary frequency $\alpha(i\omega)$ of alkali-metal and alkaline-earth-metal atoms are taken from Ref.~\cite{DerevienkoADNDT10}, whereas the dynamic polarizabilities of benzene, naphthalene, and azulene are obtained by using the explicitly connected representation of the expectation value and polarization propagator within the coupled cluster method~\cite{MoszynskiCCCC05,KoronaMP06}. 

All electronic structure calculations are performed with the \textsc{Molpro} package of \textit{ab initio} programs \cite{Molpro1,MOLPRO-WIREs}, while the force field optimizations are carried out with the \textsc{Mathematica} program~\cite{Mathematica}.

\section{Numerical results and discussion}
\label{sec:results}

\subsection{Benzene}

\begin{figure*}[t!]
\begin{center}
\includegraphics[width=0.95\textwidth]{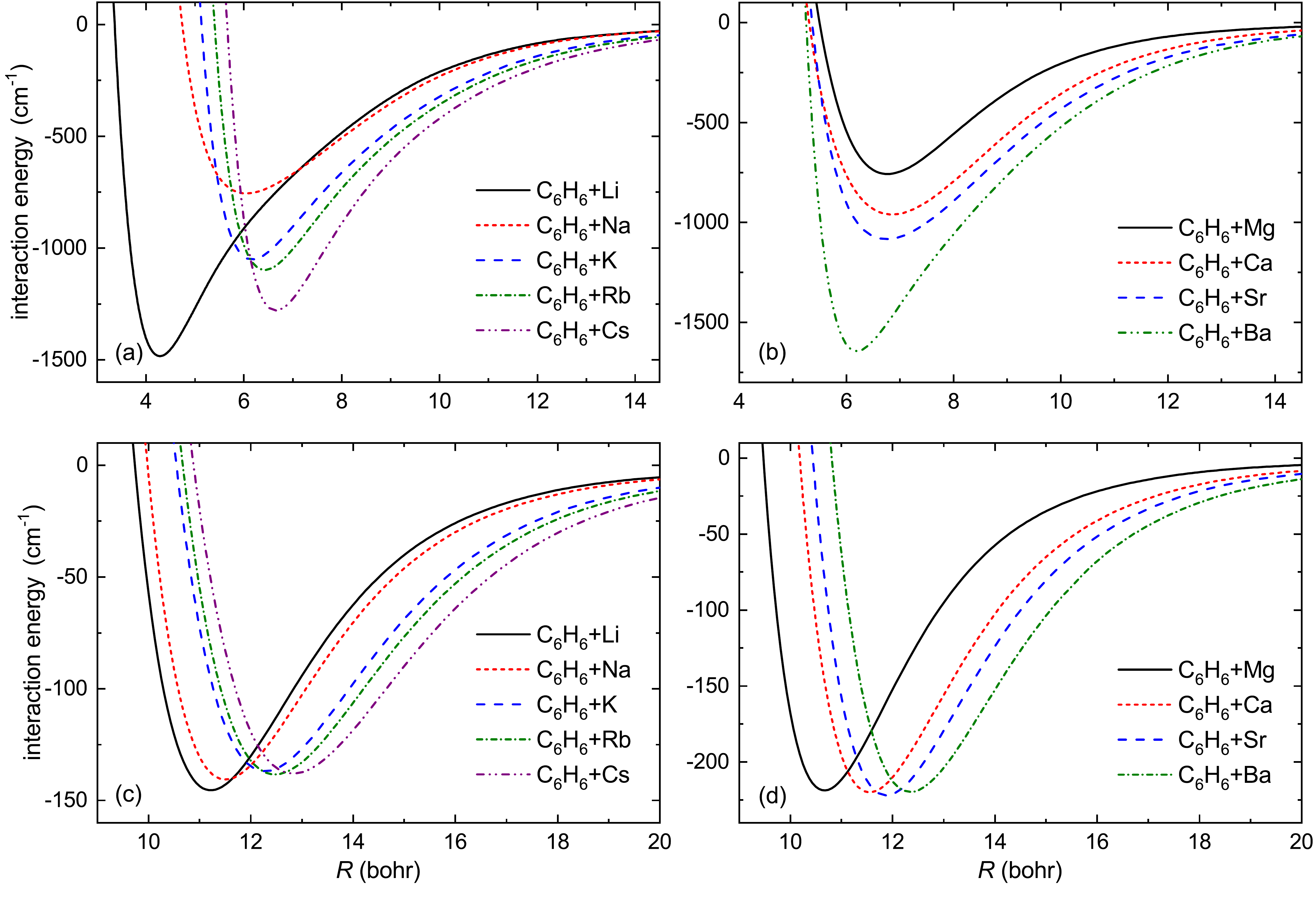}
\end{center}
\caption{One-dimensional cuts through the ground-state PESs of benzene interacting with alkali-metal (a,c) and alkaline-earth-metal (b,c) atoms at the out-of-plane (a,b) and side-in-plane (c,d) geometries obtained with the CCSD(T) method.}
\label{fig:1dPES}
\end{figure*}

The interaction energies between benzene and metal atoms are investigated for three geometries: out-of-plane $(0,0)$, side-in-plane $(\frac{\pi}{2},\frac{\pi}{6})$, and vertex-in-plane $(\frac{\pi}{2},0)$, where $(\theta,\phi)$ are the polar and azimuthal angles as introduced in Fig.~\ref{fig:geom}.
One-dimensional cuts through the ground-state PESs of benzene interacting with the Li, Na, K, Rb, Cs alkali-metal and Mg, Ca, Sr,
Ba alkaline-earth-metal atoms at out-of-plane and side-in-plane geometries are presented in Fig.~\ref{fig:1dPES}. The equilibrium intermolecular distances $R_e$ and well depths $D_e$ corresponding to the three considered arrangements are collected in Table~\ref{tab:spec}. 

At the out-of-plane geometry, when a metal atom approaches and interacts with the cloud of $\pi$ electrons, the interaction energy is an order of magnitude larger than at the side-in-plane geometry. At the out-of-plane geometry the PES well depths are between 756$\,$cm$^{-1}$ for C$_6$H$_6$+Na and 1640$\,$cm$^{-1}$ for C$_6$H$_6$+Ba. Interestingly, for all alkali-metal atoms except lithium and all alkaline-earth-metal atoms, the well depth increases systematically with the size and polarizability of a metal atom. This suggests that the interaction at the out-of-plane geometry is of the dispersion-dominated van der Waals nature. Additionally, the equilibrium intermolecular distance increases for alkali-metal atoms from 4.25$\,$bohr for C$_6$H$_6$+Li to 6.65$\,$bohr for C$_6$H$_6$+Cs and decreases for alkaline-earth-metal atoms from  6.86$\,$bohr for C$_6$H$_6$+Ca to 6.20$\,$bohr for C$_6$H$_6$+Ba with the increasing size of an atom. The much stronger and shorter-range interaction between benzene and lithium results from a larger binding energy and smaller size of the valence $s$ orbital of the Li atom, which thus favorably overlap and mix with $\pi$ electrons of benzene. Such a stronger interaction may distort the planar structure of benzene, which may additionally increase the interaction energy. In fact, such a behavior associated with a charge-transfer from the Li atom to benzene and a reduction of the system's symmetry from $C_{6v}$ to $C_{2v}$ was theoretically predicted~\cite{BakerJPCA10,DenisCPL13,SadlejPCCP15,BorcaJPCA16}.

At the side-in-plane and vertex-in-plane geometries, when a metal atom approaches and interacts with hydrogen atoms, there exist saddle points of the PESs with the interaction energy in the range of 100-200$\,$cm$^{-1}$. Interestingly, there is no dependence of the well depth on an involved atom for both alkali-metal and alkaline-earth-metal atoms, however the equilibrium intermolecular distance increases with the size of a metal atom. At the side-in-plane geometry the well depth is around 140$\,$cm$^{-1}$ and 220$\,$cm$^{-1}$ for alkali-metal and alkaline-earth-metal atoms, whereas at the vertex-in-plane  geometry the well depth is around 110$\,$cm$^{-1}$ and 160$\,$cm$^{-1}$ for alkali-metal and alkaline-earth-metal atoms, respectively. At the side-in-plane geometry, the equilibrium intermolecular distance increases from 11.2$\,$bohr for C$_6$H$_6$+Li to 12.8$\,$bohr for C$_6$H$_6$+Cs and  from  10.7$\,$bohr for C$_6$H$_6$+Mg to 12.3$\,$bohr for C$_6$H$_6$+Ba with the increasing size of an atom. At the vertex-in-plane geometry, the equilibrium intermolecular distance increases from 11.9$\,$bohr for C$_6$H$_6$+Li to 13.4$\,$bohr for C$_6$H$_6$+Cs and  from  11.5$\,$bohr for C$_6$H$_6$+Mg to 13.0$\,$bohr for C$_6$H$_6$+Ba with the increasing size of an atom.

\begin{figure*}[tbh!]
\begin{center}
\includegraphics[width=0.95\textwidth]{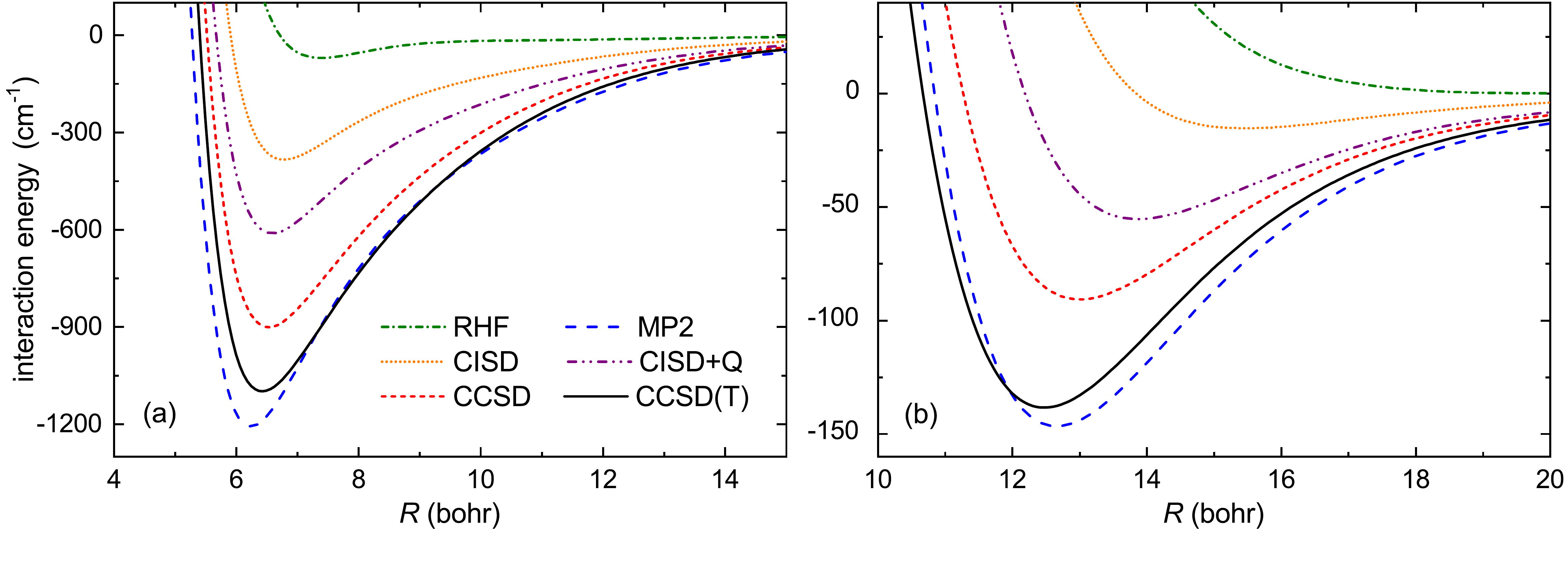}
\end{center}
\caption{One-dimensional cuts through the ground-state PES of benzene interacting with rubidium atom at the out-of-plane (a) and side-in-plane (b) geometries calculated at the RHF, MP2, CISD, CISD+Q, CCSD, and CCSD(T) levels of theory.}
\label{fig:methods}
\end{figure*}

To evaluate the performance of the used \textit{ab initio} methods in the reproduction of the correlation energy, in Fig.~\ref{fig:methods} we present one-dimensional cuts through the ground-state PES of benzene interacting with rubidium atom at the out-of-plane and side-in-plane geometries obtained at the RHF, MP2, CISD, CISD+Q, CCSD, and CCSD(T) levels of theory. As expected, there is no stabilizing interaction at the mean-field level of the restricted Hartree-Fock (RHF) calculations with a purely repulsive potential at the side-in-plane geometry. The configuration interaction method including single and double excitations (CISD) poorly reproduces correlation energy, however correctly locates the equilibrium distance at the out-of-plane geometry. The inclusion of the Davidson correction to the configuration interaction results (CISD+Q) improves the description, however the interaction energy is still underestimated by 50\%. The coupled cluster method including single and double excitations (CCSD) significantly outperforms the configuration interaction method. A large discrepancy between the CCSD and CISD (note parenthetically that the latter method is size-inconsistent) indicates a significant contribution from the interaction between electron-correlated parts of the respective monomer wave functions (e.g.~terms which stem from simultaneously doubly-excited configurations on benzene and rubidium). Further inclusion of non-iterative triple excitations in the coupled cluster method (CCSD(T)) accounts for around 30\% of the total interaction energy. Interestingly, the second-order M{\o}ller-Plesset perturbation theory (MP2) slightly overestimates the interaction energy as compared to the CCSD(T) method but outperforms the CCSD method. The above observation is a result of an accidental error cancellation but it suggests that the use of the MP2 method can be a reasonable choice for generating PESs or optimizing equilibrium geometries for systems of polyatomic aromatic molecules interacting with metal atoms. It should be noted, however, that the utilization of MP2 as a supermolecular method for the intermolecular interactions between complexes containing stacked $\pi$ systems is discouraged, as it has a tendency to significantly overestimate the binding energy~\cite{Hobza:96}.

Based on the above considerations and additional analysis of the convergence with the size of the used atomic basis sets and performance of the employed set of the bond functions, we estimate that the uncertainty of calculated PESs is of the order of 10-20\%. Most probably we underestimate the interaction energy. Our results also agree within our estimated error bars with recent calculations for benzene interacting with selected metal atoms~\cite{DenisCPL13,DenisCP14,UllahCPL18}.

\begin{figure*}[t!]
\begin{center}
\includegraphics[width=0.95\textwidth]{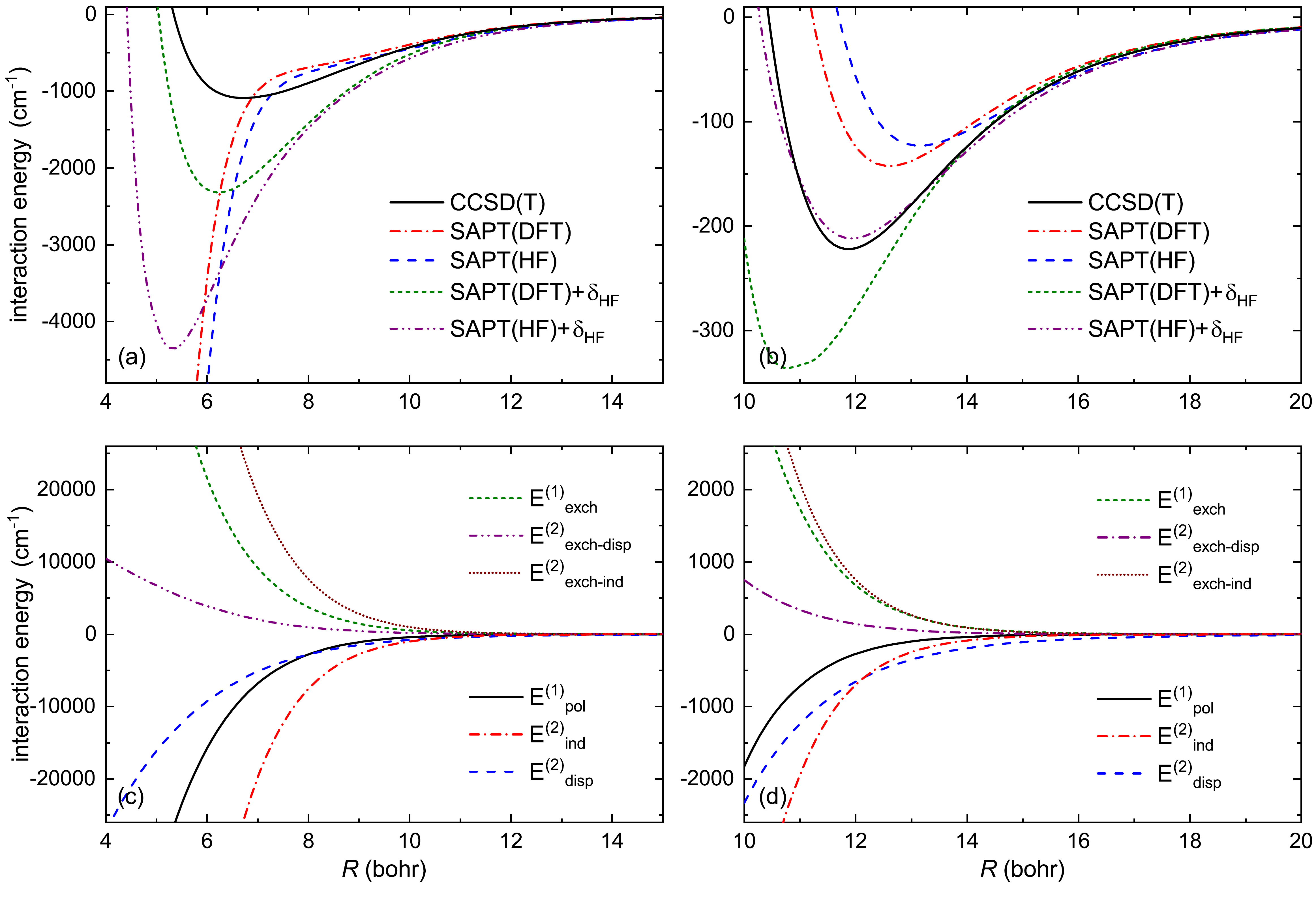}
\end{center}
\caption{(a),(b) One-dimensional cuts through the ground-state PES of benzene interacting with strontium atom obtained with different variants of the SAPT approach and compared with the CCSD(T) results. (c),(d) Decomposition of the interaction energy into SAPT(HF) components. Results are presented for the out-of-plane (a),(c) and side-in-plane (b),(d) geometries.}
\label{fig:sapt}
\end{figure*}

\begin{figure*}[t!]
\begin{center}
\includegraphics[width=0.95\textwidth]{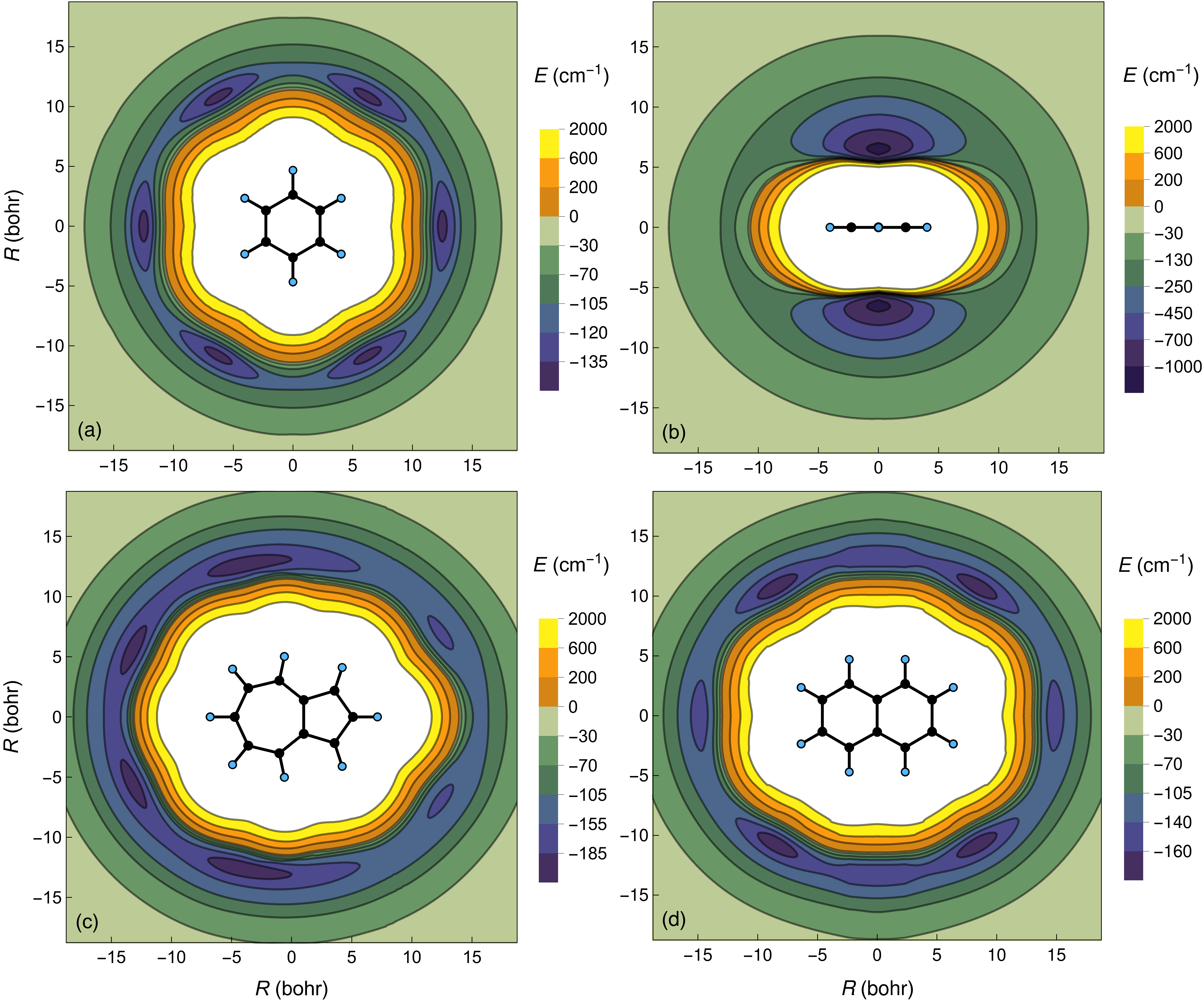}
\end{center}
\caption{Two-dimensional cuts through the ground-state PESs of benzene interacting with rubidium atom at the in-plane (a) and out-of-plane (b) geometries and of azulene (c) and naphthalene (d) interacting with rubidium atom at the in-plane geometry obtained with the CCSD(T) method. Note different energy scales for different panels.}
\label{fig:surfs}
\end{figure*}

The SAPT calculations for the examplary system of a complex of benzene and strontium were performed for selected one-dimensional cuts through the PES
in order to examine the applicability of this theory for the computation of the interaction energy between an aromatic molecule and an metal atom and to identify the main components of the intermolecular interaction energy. The results, presented in Fig.~\ref{fig:sapt}, show that
the long-range behavior is dominated, as expected, by the second-order dispersion term, which decays with the sixth inverse power of the intermolecular distance, since the long-range induction decays faster than the dispersion contribution. The first-order electrostatic term is negligible for large distances, since its behavior is short-range if one of interacting species is an atom. Summarizing, the long-range SAPT interaction energy agrees quite well with the CCSD(T) benchmark results independently of the geometry type (out-of-plane or side-in-plane).

A completely different picture arises at distances closer to the PES minimum. One can see a growing discrepancy between the SAPT and CCSD(T) interaction energies. For the out-of-plane geometry both SAPT(HF) and SAPT(DFT) curves are several times deeper than CCSD(T) one and the very existence of their minima is dependent on an addition of the $\delta E_\text{HF}$ term (which in principle should be a small correction with respect to other SAPT components). A large absolute value of this term indicates that the finite-order SAPT has serious problems with recovering accurate values of the interaction energy. The out-of-plane SAPT(DFT) minimum lies about 0.75\,bohr closer and is two times deeper than the CCSD(T) benchmark values,
while the situation for the SAPT(HF) is even worse: its depth is four times too big in comparison to the benchmark. Several explanations to this behavior can be found in the SAPT literature. The first one seeks for the problem in a simplified treatment of exchange SAPT terms, which
are calculated in the so-called single-exchange ($S^2$) approximation, i.e.~which skips multiple exchanges of electrons between the monomers,
while another explanation assumes that in some cases one can encounter the ``polarization catastrophy" phenomenon, which may occur because
the Pauli exclusion principle is not enforced on level of wave functions in a so-called weak symmetry-forcing employed in symmetrized Rayleigh-Schr{\"o}dinger (SRS) perturbation theory~\cite{Jeziorski:78}. For instance, the neglection of terms higher than $S^2$ was identified as a main culprit of a poor behavior of SAPT employed for the calculation of metal dimers' interaction energies~\cite{Patkowski2007}, where a simple rescaling of the second-order exchange corrections by the $E^{(1)}_\text{exch}/E^{(1)}_\text{exch}(S^2)$ ratio has been proposed as a partial remedy. 
An approach allowing for avoiding the $S^2$ approximation within single-determinant SAPT and a pilot implementation for small molecules has been reported in Ref.~\cite{Schaeffer2012,Schaeffer2013} some time ago. Some trials of treatment of the second problem have been reported e.g.~in Ref.~\cite{Patkowski2004} through the use of the regularized potential, but they never reached the mature stage and we cannot utilize them in our case.

A detailed analysis of SAPT energy components reveals indeed a very large absolute value of the second-order induction energy, especially its benzene$\rightarrow$Sr component, which is only partly compensated by the corresponding exchange-induction term. It can be also seen that a difference between the $E^{(1)}_\text{exch}$ term (calculated without the $S^2$ approximation) and $E^{(1)}_\text{exch}(S^2)$ is quite significant at distances close to the minimum (e.g.~for $R=6.5$\,bohr for the out-of-plane geometry it amounts to 733\,cm$^{-1}$ for SAPT(HF), which should be compared with $-1076$\,cm$^{-1}$ of CCSD(T) interaction energy for this distance). However, an approximation of the second-order exchange terms by utilizing the same ratio as in Ref.~\cite{Patkowski2007} leads to a huge overestimation of the resulting interaction energy. We can therefore conclude that the breakdown of the $S^2$ approximation can be responsible for the failure of the SAPT in recovering the interaction energy for this system. 
This hypothesis is supported by results from Ref.~\cite{Schaeffer2013}, where the underestimation of the first-order exchange and second-order exchange-induction, and overestimation of the exchange-dispersion terms 
calculated within the $S^2$ approximation has been numerically detected for complexes like Ar$_2$ or (H$_2$O)$_2$. Since Figure~\ref{fig:sapt} shows a much higher importance of the exchange-induction term
in comparison to the exchange-dispersion
in our case, it can be anticipated that the increased repulsion coming from missing multiple-exchange terms could reduce the gap between the SAPT and benchmark CCSD(T) results.
On the other hand the large absolute value of the second-order induction energy points to the ``polarization catastrophy" phenomenon as another culprit for this failure.
The second cause is especially probable because  of a large polarizability of the valence electrons of the alkaline-earth atoms which come into
an easy interference with the loosely bound $\pi$ electrons of the benzene ring. Therefore, one can conclude that unfortunately the SAPT, both in the HF and DFT flavors, should not be utilized in this case as a quantitative model.

The same conclusions can be reached when analysing the second orientation. Also in this case the $\delta E_\text{HF}$ term is indispensable instead of being a small correction. For SAPT(DFT) the minimum is placed about 1\,bohr closer and is about one-half deeper than the benchmark CCSD(T) one and without the $\delta E_\text{HF}$ it would be shallower than CCSD(T) and shifted by 1.5\,bohr towards larger distances. Probably accidentally, the SAPT(HF) value (with delta Hartree-Fock) is close to CCSD(T) -- it should be noted that for this geometry a similar situation occurs for the supermolecular MP2 energy, which is closely related to SAPT(HF)~\cite{Chalasinski:88}.

The leading long-range dispersion interaction coefficients are reported in Table~\ref{tab:Cn_benzen}. Their values indicate moderate anisotropy of the long-range interaction potential.

\begin{table*}[tbh!]
\caption{Parameter values of the used force field model describing interactions in benzene--metal-atom systems fitted to the present \textit{ab initio} data. \label{tab:forcefield}} 
\begin{ruledtabular}
\begin{tabular}{lrrrrrrrr}
System & 
$r_\mathrm{CC}^{\parallel}$ (bohr) & $\epsilon_\mathrm{CC}^{\parallel}$ (cm$^{-1}$) &   $r_\mathrm{CC}^{\perp}$ (bohr) & $\epsilon_\mathrm{CC}^{\perp}$ (cm$^{-1}$)&   $r_\mathrm{CH}^{\parallel}$ (bohr) & $\epsilon_\mathrm{CH}^{\parallel}$ (cm$^{-1}$)&   $r_\mathrm{CH}^{\perp}$ (bohr) & $\epsilon_\mathrm{CH}^{\perp}$ (cm$^{-1}$)\\
\hline
C$_6$H$_6$+Li & 4.49 & 118 & 11.9 & 19.0 & 6.03 & 177 & 5.24 & 153\\
C$_6$H$_6$+Na &	5.45 & 45.1 & 12.4 & 14.2 & 7.17 & 122 & 5.70 & 72.7\\
C$_6$H$_6$+K & 6.74 & 35.2 & 13.2 & 15.7 & 7.13 & 181 & 5.71 & 40.6\\
C$_6$H$_6$+Rb & 6.80 & 62.2 & 13.1 & 17.9 & 7.45 & 148 & 5.07 & 144\\
C$_6$H$_6$+Cs & 7.17 & 133 & 13.4 & 19.8 & 7.21 & 106 & 4.26 & 109\\
C$_6$H$_6$+Mg & 5.94 & 81.8 & 10.6 & 11.2 & 8.0 & 94.5 & 3.93 & 123 \\
C$_6$H$_6$+Ca & 5.14 & 99.2 & 12.4 & 20.0 & 7.75 & 160 & 4.17 & 148\\
C$_6$H$_6$+Sr & 5.67 & 100 & 12.7 & 23.1 & 7.78 & 165 & 4.53 & 150\\
C$_6$H$_6$+Ba & 6.36 & 127 & 13.6 & 20.8 & 7.57 & 186 & 6.78 & 87.8 
\end{tabular}
\end{ruledtabular}
\end{table*}

\begin{figure*}[tbh!]
\begin{center}
\includegraphics[width=0.95\textwidth]{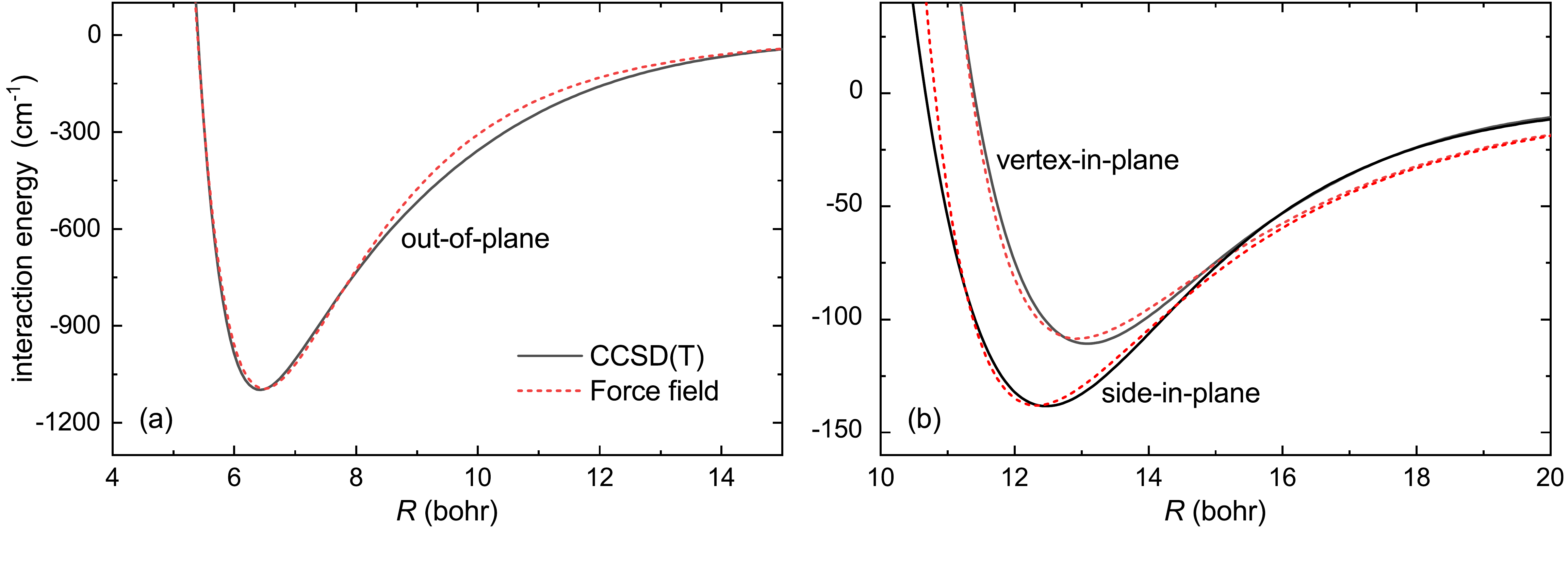}
\end{center}
\caption{One-dimensional cuts through the ground-state PES of benzene interacting with rubidium at the out-of-plane (a) and side-in-plane and vertex-in-plane (b) geometries obtained with the optimized force field and compared with the CCSD(T) results.}
\label{fig:FFvsCCSDT}
\end{figure*}

\subsection{3D PES and force field}

Two-dimensional cuts through the ground-state PES of benzene interacting with rubidium atom at the in-plane and out-of-plane geometries are presented in Fig.~\ref{fig:surfs}(a) and Fig.~\ref{fig:surfs}(b). Two global minima at the out-of-plane axis are clearly visible. Additionally, two sets of equivalent saddle points of the $C_{6v}$ symmetry are pronounced at the in-plane geometry. The PES at small intermolecular distances is strongly anisotropic. Interestingly, the PES at an intermolecular distance of $15$~bohr or more starts to be close to spherically symmetric, in agreement with moderate values of the anisotropic long-range dispersion coefficients. This indicates that benzene in cold and ultracold collisions may behave as a relatively spherical molecule with a large and favorable ratio of elastic to rotationally inelastic cross sections.

Both quantum and classical scattering calculations need reliable PESs as an input. A generation of accurate three-dimensional PESs for rigid polycyclic aromatic hydrocarbon molecules and their substituted derivatives interacting with metal atoms is computationally very challenging. A parameterization of such interactions by a force field may be a remedy. 

Here, we use the force field within the atom-bond pairwise additive representation suggested in Ref.~\cite{PiraniAtomBond} and slightly modified as described in Sec.~\ref{sec:theory}. The modification was necessary to add more flexibility to the model to account for strongly anisotropic short-range interactions. Our force field model given by Eqs.~\eqref{eq:lk}-\eqref{eq:ff} is fitted to the present \textit{ab initio} data, that is series of points calculated at out-of-plane, side-in-plane, and vertex-in-plane geometries. Parameter values of the optimized force field model describing interactions between benzene and metal atoms are collected in Table~\ref{tab:forcefield}. The performance of the optimized force field is evaluated by comparison with the CCSD(T) results in Fig.~\ref{fig:FFvsCCSDT} for the exemplary benzene-rubidium system. For other metal atoms very similar agreement is obtained.  The characteristics of the studied systems, presented in Table~\ref{tab:spec}, are reproduced by our force field on average within 1.5\% in the value of equilibrium distances and 3\% in the value of well depths. Unfortunately, slightly worse performance is observed for larger distances and configurations between in-plane and out-of-plane geometries. Nevertheless, our force fields can be used in scattering calculations for collisions between benzene and metal atoms, as well as can give useful information about interactions between substituted derivatives of benzene and polycyclic aromatic hydrocarbons with metal atoms. The transferability of the proposed force field between different aromatic systems, however, still has to be verified.

\subsection{Naphthalene and azulene}

\begin{table}[t!]
\caption{Properties of benzene, naphthalene, and azulene at the equilibrium geometry: cartesian components of the static electric dipole polarizability $\alpha_e^{ii}$ (in atomic units) and permanent electric dipole moment $d_e$ (in Debye) obtained with the the CCSD(T) method within the finite field approach and aug-cc-pVTZ basis sets. Equilibrium interatomic distances or their ranges ($r_e^\text{CC}$, $r_e^\text{CH}$ in atomic units) are presented as determined by high-resolution spectroscopy~\cite{BabaJCP11,HuberMP05}.\label{tab:molecules}} 
\begin{ruledtabular}
\begin{tabular}{lrrrrrrr}
Molecule & $\alpha_e^{xx}\,$ & $\alpha_e^{yy}\,$ & $\alpha_e^{zz}\,$  & $|d_e|\,$ & $r_e^\text{CC}$ & $r_e^\text{CH}$ \\
\hline
benzene     &  79.0 & 79.0 &44.1 & - & 1.40 & 1.08 \\
naphthalene &  165 & 122 & 65.8 & - & 1.38-1.43 & 1.08\\
azulene     &  191 & 129 & 67.2 &  0.947 & 1.38-1.48 & 1.08 \\
\end{tabular}
\label{tab:prop}
\end{ruledtabular}
\end{table}

\begin{table}[b!]
\caption{Characteristics of the PESs for naphthalene interacting with selected alkali-metal and alkaline-earth-metal atoms, all in the ground electronic state: equilibrium intermolecular distance $R_e$ and well depth $D_e$ for saddle points at the main axes of the in-plane geometry.} 
\begin{ruledtabular}
\begin{tabular}{lrrrr}
System & $R_e\,$(bohr) & $D_e\,$(cm$^{-1}$) & $R_e'\,$(bohr) & $D_e'\,$(cm$^{-1}$) \\
\hline
C$_{10}$H$_8$+Li & 13.6 & 157 & 11.9 & 163 \\
C$_{10}$H$_8$+Rb & 14.8 & 152 & 13.2 & 160  \\
C$_{10}$H$_8$+Sr & 14.2 & 236 & 12.6 & 241
\end{tabular}
\label{tab:spec_nph}
\end{ruledtabular}
\end{table}

Naphthalene and its isomer azulene are the simplest polycyclic aromatic hydrocarbon molecules. Some of their properties are compared with those of benzene in Table~\ref{tab:prop}. Two-dimensional cuts through the ground-state PESs of these molecules interacting with rubidium atom at the in-plane geometry are presented in Fig.~\ref{fig:surfs}(c) and Fig.~\ref{fig:surfs}(d) and can be compared with those of benzene in Fig.~\ref{fig:surfs}(a). Two sets of eight saddle points of the $C_{2v}$ symmetry are pronounced for naphthalene and two sets of six saddle points of the $C_{s}$ symmetry are visible for azulene. The equilibrium intermolecular distances $R_e$ and well depths $D_e$ of the PESs for saddle points of naphthalene interacting with selected alkali-metal and alkaline-earth-metal atoms at the main axes of the in-plane geometry are collected in Table~\ref{tab:spec_nph}. 

\begin{figure}[tbt!]
\begin{center}
\includegraphics[width=0.95\columnwidth]{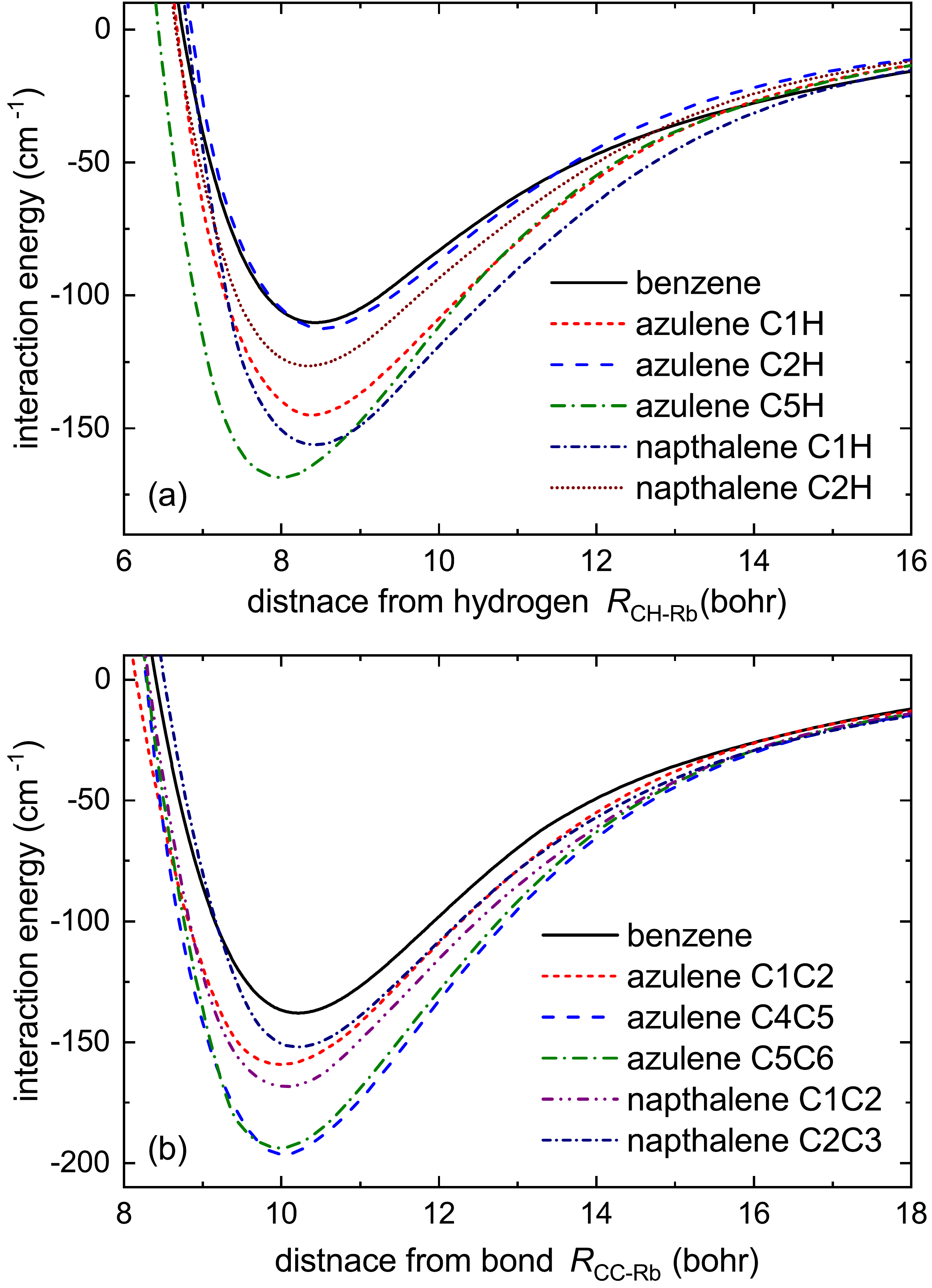}
\end{center}
\begin{center}
\includegraphics[width=0.45\columnwidth]{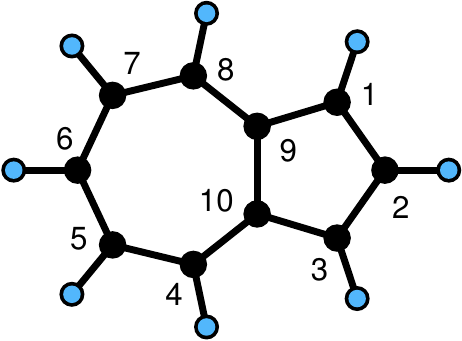}
\includegraphics[width=0.40\columnwidth]{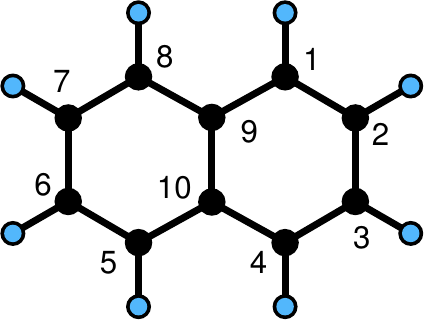}
\end{center}
\caption{One-dimensional cuts through the ground-state PESs of benzene, naphthalene, and azulene interacting with rubidium at sevral vertex-in-plane (a) and side-in-plane (b) geometries obtained with the CCSD(T) method. Used labeling of carbon atoms is presented in the bottom panel.}
\label{fig:AvsB}
\end{figure}

The global minima of PESs for naphthalene and azulene interacting with metal atoms lie at the out-of-plane geometries, symmetrically above and below the clouds of delocalized $\pi$ electrons. Unfortunately, our computational method is not able to reproduce smoothly the interaction energy for the out-of-plane geometries, because the energies of low lying exited states of the considered molecule-atom systems approach the energies of the ground state at small intermolecular distances resulting in the multireference character of the systems. The leading CI coefficients at the MCSCF level always exceed 0.99 for all geometries of benzene and side-in-plane geometries of naphthalene and azulene, but reach $1/\sqrt{2}$ for out-of-plane geometries of naphthalene and azulene, when the excited to $p$ orbital electron of alkali-metal atom is strongly stabilized by the interaction with $\pi$ electrons of polycyclic aromatic molecule.  Nevertheless, our calculations suggest that for naphthalene interacting with metal atoms there are two sets of two equivalent global minima in the form of shallow double wells similarly as it was predicted for naphthalene interacting with noble-gas atoms~\cite{ClementiJPCA01,MakarewiczJCP11}. For azulene interacting with metal atoms, the interaction energy at global minima tends to be much larger as compered to benzene and naphthalene, probably because of a charge separation and dipole moment in azulene related to an electron transferred from its seven membered ring (making it a tropylium cation) into its five membered ring (making it a cyclopentadienyl anion).
 
Similarly as for benzene, the PESs for naphthalene and azulene interacting with metal atoms at small intermolecular distances are strongly anisotropic, however they start to be closer to spherically symmetric at  intermolecular distances of $15$~bohr or more. This indicates that naphthalene and azulene in cold and ultracold collisions may behave as relatively spherical molecules with large and favorable ratios of elastic to rotationally inelastic cross sections. The corresponding leading long-range dispersion and induction interaction coefficients are reported in Table~\ref{tab:Cn_benzen} and Table~\ref{tab:Cn_azulene}. Their values indicate moderate anisotropy of the long-range interaction potential, which however is twice larger as compared to benzene.

To compare the interactions in the investigated systems, in Fig.~\ref{fig:AvsB} we plot one-dimensional cuts through the ground-state PESs of benzene, naphthalene, and azulene interacting with rubidium at several in-plane geometries. Interestingly, if we parametrize vertex-in-plan geometries by the distance between metal and hydrogen atoms, and side-in-plane geometries by the distance between a metal atoms and a center of bond, the corresponding curves have similar characteristics. The equilibrium distances agree within a few percent, whereas the well depths for naphthalene and azulene are not more than by 50\% larger as compared with benzene. Specifically, the largest and smallest interaction energies are for rubidium atom interacting with the bigger and smaller rings of azulene, while interaction energies for naphthalene lie between. The order of curves for naphthalene can be explained by the steric effects, whereas for azulene the interplay of its electric dipole moment, steric effects, and electron density distribution is responsible for observed differences.    

The universality of interactions between studied systems visible in Fig.~\ref{fig:AvsB} raises the question about the ability of our force field developed for benzene--metal-atom systems to reproduce the interaction energies for polycyclic aromatic hydrocarbon molecules interacting with metal atoms. Unfortunately, the performance of the proposed force field with parameters reported in Table~\ref{tab:forcefield} is not fully satisfactory leading to under- or overestimation of interactions energies by up to 50\%. Possible explanation of this failure may lie in the form of the used force field model which was initially developed to describe weaker interaction between benzene and noble-gas atoms~\cite{PiraniAtomBond}, on one hand, and the charge separation and dipole moment in the case of azulene, on the other hand. Further studies on transferable force fields for polycyclic aromatic hydrocarbon molecules are thus needed. Nevertheless, the present force field can be a reasonable starting point for investigating collisional dynamics of polycyclic aromatic molecules with metal atoms.

\begin{table}[tb!]
\caption{Isotropic and anisotropic dispersion coefficients describing the long-range part of the interaction potential of benzene and naphthalene interacting with alkali-metal and alkaline-earth-metal atoms.\label{tab:Cn_benzen}} 
\begin{ruledtabular}
\begin{tabular}{lrr}
System &   $C^{\mathrm{disp}}_{6,0}\,$(a.u.) & $C^{\mathrm{disp}}_{6,2}\,$(a.u.)\\
\hline
C$_6$H$_6$+Li & 1044 & -176\\
C$_6$H$_6$+Na &	1167 & -195\\
C$_6$H$_6$+K & 	1765 & -293\\
C$_6$H$_6$+Rb & 1979 & -327\\
C$_6$H$_6$+Cs & 2389 & -393\\
C$_6$H$_6$+Mg & 922& -148\\
C$_6$H$_6$+Ca & 1571 & -255\\
C$_6$H$_6$+Sr & 1880 & -306\\
C$_6$H$_6$+Ba & 2361 & -384\\
\hline
C$_{10}$H$_8$+Li & 1809 &-385\\
C$_{10}$H$_8$+Na & 2018 &-424\\
C$_{10}$H$_8$+K  & 3051 &-640\\
C$_{10}$H$_8$+Rb & 3417 &-711\\
C$_{10}$H$_8$+Cs & 4123 &-854\\
C$_{10}$H$_8$+Mg & 1576 &-316\\
C$_{10}$H$_8$+Ca & 2698 &-550\\
C$_{10}$H$_8$+Sr & 3231 &-659\\
C$_{10}$H$_8$+Ba & 4059 &-830\\
\end{tabular}
\end{ruledtabular}
\end{table}
\begin{table}[tb!]
\caption{Isotropic and anisotropic dispersion and induction coefficients describing the long-range part of the interaction potential of azulene interacting with alkali-metal and alkaline-earth-metal atoms.\label{tab:Cn_azulene}} 
\begin{ruledtabular}
\begin{tabular}{lrrrr}
System &   $C^{\mathrm{disp}}_{6,0}\,$(a.u.) & $C^{\mathrm{ind}}_{6,0}\,$(a.u.) & $C^{\mathrm{disp}}_{6,2}\,$(a.u.) & $C^{\mathrm{ind}}_{6,2}\,$(a.u.)\\
\hline
C$_{10}$H$_8$+Li & 1996 & 22.8 & -467 & 22.8 \\
C$_{10}$H$_8$+Na & 2220 & 23.1 & -513 & 23.1 \\
C$_{10}$H$_8$+K  & 3360 & 40.4 & -776 & 40.4 \\
C$_{10}$H$_8$+Rb & 3756 & 44.4 & -861 & 44.4 \\
C$_{10}$H$_8$+Cs & 4529 & 54.9 & -1033 & 54.9 \\
C$_{10}$H$_8$+Mg & 1709 & 10.0 & -374 & 10.0 \\
C$_{10}$H$_8$+Ca & 2942 & 21.8 & -657 & 21.8 \\
C$_{10}$H$_8$+Sr & 3525 & 27.7 & -788 & 27.7 \\
C$_{10}$H$_8$+Ba & 4435 & 38.4 & -995 & 38.4
\end{tabular}
\end{ruledtabular}
\end{table}

Among the considered polyatomic molecules, azulene may be especially interesting for cold studies. On one hand, a significant permanent electric dipole moment of almost one Debye may be useful for electric field manipulation relevant for guiding and cooling techniques such as Stark~\cite{BethlemNature00} and centrifuge~\cite{WuScience17} decelerating or Sisyphus laser cooling~\cite{KozyryevPRL17a}. On the other hand, it possesses optical transitions in the visible range~\cite{FabriziaJCP93,BearparkJACS96,SembaJCP09}, which may potentially be useful for laser manipulation and high precision spectroscopy, which can reveal subtle details of rovibrational dynamics and rovibranic couplings in polyatomic molecules. Studies on collisional intermolecular energy transfer between vibrationally excited azulene and noble-gas atoms have already been started~\cite{LimJPC90,BernshteinJCP06,BernshteinMP08,KimJPCA16}. Another polyatomic aromatic hydrocarbon molecules that possess permanent electric dipole moment is fulvene~\cite{BaronJMS72}, an isomer of benzene. It, however, possess less favorable optical transitions, therefore we have selected azulene for the present work.

\section{SUMMARY AND CONCLUSIONS}
\label{sec:summary}

Motivated by recent interest and advances in cooling and application of polyatomic molecules at low and potentially ultralow temperatures, we have  considered collisional properties of benzene, naphthalene, and azulene immersed into ultracold gases of alkali-metal and alkaline-earth-metal atoms. To this end we have calculated and characterized potential energy surfaces and leading long-range interaction coefficients in these systems by using state-of-the-art ab initio techniques: the coupled cluster method restricted to single, double, and noniterative triple excitations, CCSD(T), combined with large Gaussian basis sets and small-core energy-consistent pseudopotentials. We have analyzed and benchmarked the nature of intermolecular interactions using symmetry-adapted perturbation theory, which unfortunately fails in recovering accurately the interaction energy. We have also pointed the need for the multireference description of out-of-plane interactions of $\pi$ electrons in polycyclic aromatic molecules with polarizable metal atoms. We have provided the full three-dimensional PESs for selected systems within the atom-bond pairwise additive representation. We have suggested azulene, an isomer of naphthalene which possesses a significant permanent electric dipole moment and optical transitions in the visible range, as a promising candidate for electric field manipulation and buffer-gas or sympathetic cooling. A relatively weak anisotropy of long-range interactions in the investigated systems may result in favorable ratios of elastic to rotationally inelastic cross sections, and suggests good prospects for collisional cooling.

The present study of the intermolecular interactions is the first step towards the evaluation of prospects for sympathetic cooling and controlled chemistry of polyatomic aromatic molecules with ultracold alkali-metal or alkaline-earth-metal atoms. This work also establishes and benchmarks the computational scheme for the future \textit{ab initio} investigations of intermolecular interactions in other polyatomic aromatic molecule-atom systems relevant for ultracold physics and chemistry. In the future, the obtained PESs and long-range interaction coefficients will be employed in time-independent scattering calculations for both elastic and inelastic collisions at low and ultralow temperatures to evaluate prospects for sympathetic cooling. The present results may also be useful for better understanding interactions between polycyclic aromatic hydrocarbons and metal atoms in lithium batteries, hydrogen storage devices, and alkali-metal-doped carbon-based superconductors~\cite{KubozonoPCCP11}.

\begin{acknowledgments}
Financial support from the National Science Centre Poland (2015/19/D/ST4/02173), the Foundation for Polish Science within the Homing programme co-financed by the European Union under the European Regional Development Fund, and the PL-Grid Infrastructure is gratefully acknowledged. 
\end{acknowledgments}

\bibliography{benzen}

\end{document}